\documentclass[12pt,epsf]{article}
\usepackage[dvips]{graphicx}

\baselineskip=\normalbaselineskip

\renewcommand{\thefootnote}{\fnsymbol{footnote}}
\newcommand{\bel}[1]{\begin{equation}\label{#1}}
\newcommand{\bal}[1]{\begin{eqnarray}\label{#1}}

\newcommand{\be}{\begin{equation}}
\newcommand{\ee}{\end{equation}}
\newcommand{\ba}{\begin{eqnarray}}
\newcommand{\ea}{\end{eqnarray}}
\newcommand{\nn}{\nonumber \\}

\newcommand{\bR}{{\bf R}}

\newcommand{\eq}[1]{(\ref{#1})}

\newcommand{\cL}{{\cal L}}

\newcommand{\Sloc}{{S_{\rm loc}}}
\newcommand{\cLloc}{{{\cal L}_{\rm loc}}}
\newcommand{\cJ}{{\cal J}}

\newcommand{\cW}{{\cal W}}

\newcommand{\Gb}{\overline{G}}
\newcommand{\phib}{\bar{\phi}}

\newcommand{\bG}{\mbox{\boldmath $G$}}
\renewcommand{\bR}{\mbox{\boldmath $R$}}
\ifx\TwoupWrites\UnDeFiNeD\else\target{\magstepminus1}{11.3in}{8.27in}
\source{\magstep0}{7.5in}{11.69in}\fi
\newfont{\fourteencp}{cmcsc10 scaled\magstep2}
\newfont{\titlefont}{cmbx10 scaled\magstep3}
\newfont{\authorfont}{cmcsc10 scaled\magstep1}
\newfont{\fourteenmib}{cmmib10 scaled\magstep2}
\skewchar\fourteenmib='177
\newfont{\elevenmib}{cmmib10 scaled\magstephalf}
\skewchar\elevenmib='177
\makeatletter
\newcommand\nonsequentialeqnum{
\@addtoreset{equation}{section}
\def\theequation{\arabic{section}.\arabic{equation}}}
\newif\ifp@bblock \p@bblocktrue
\newcommand\nopubblock{\p@bblockfalse}
\newcommand\topspace{\hrule height 0pt depth 0pt \vskip}
\newcommand\p@bblock{\begingroup \tabskip=\hsize minus \hsize
\baselineskip=1.5\ht\strutbox \topspace-2\baselineskip
\halign to\hsize{\strut ##\hfil\tabskip=0pt\crcr
\the\Pubnum\crcr\the\date\crcr}\endgroup}

\renewcommand\titlepage{\ifx\TwoupWrites\UnDeFiNeD\null
\vspace{-1.7cm}\fi
\vskip0.6cm
\ifp@bblock\p@bblock \else\hrule height 0pt \relax \fi}
\makeatother
\newtoks\date
\newtoks\Pubnum
\newtoks\pubnum
\Pubnum={
~ \crcr
~ \crcr
YITP-00-40 \crcr
{\tt hep-th/0007200} \crcr
{}July 2000
}
\newcommand{\frontpageskip}{\vspace{12pt plus .5fil minus 2pt}}
\renewcommand{\title}[1]{\frontpageskip
\begin{center}{\titlefont #1}\end{center}\par}
\renewcommand{\author}[1]{\frontpageskip\par\begin{center}
{\authorfont #1}\end{center}
\nobreak
}

\renewcommand{\thanks}[1]{\footnote{#1}}
\renewcommand{\abstract}{\par\frontpageskip\centerline{
\fourteencp Abstract}
\vspace{8pt plus 3pt minus 3pt}}

\begin{document}

\begin{titlepage}

\vspace*{\fill}
\begin{center}
{\Large{\bf Comment on Ambiguities \\
in the Holographic Weyl Anomaly}} \\
\vfill
{\sc Masafumi Fukuma
\footnote{e-mail: {\tt fukuma@yukawa.kyoto-u.ac.jp}}
and
{\sc Tadakatsu Sakai}
\footnote{e-mail: {\tt tsakai@yukawa.kyoto-u.ac.jp}}}\\[2em]

{\sl Yukawa Institute for Theoretical Physics,\\
      Kyoto University, Kyoto 606-8502, Japan } \\

\vfill
ABSTRACT
\end{center}
\begin{quote}
We consider possible ambiguities in the holographic Weyl anomaly 
that may arise from local terms in the flow equation. 
We point out that such ambiguities actually do not give 
physically meaningful contributions to the Weyl anomaly. 
\end{quote}
\vfill
\end{titlepage}

%
\renewcommand{\thefootnote}{\arabic{footnote}}
\setcounter{footnote}{0}
\addtocounter{page}{1}%

The AdS/CFT correspondence \cite{MGKPW} 
(for a review see Ref.\ \cite{review}) 
provides us with a useful framework to study the renormalization group 
(RG) structure of boundary field theories 
\cite{E.T.A}\cite{AG}\cite{HRG}\cite{GPPZ}\cite{GPPZ-2}%
\cite{PS}\cite{BPPZ}\cite{ST}\cite{DFGK}. 
In a remarkable paper \cite{dVV}, de Boer, Verlinde and Verlinde 
reformulated the holographic RG on the basis of the Hamilton-Jacobi 
equation for gravity systems, 
and introduced the ``flow equation,''
from which one can easily derive the Weyl anomaly of 
boundary conformal field theories, that is in exact agreement with the
one obtained in Ref.\ \cite{HS;weyl}. 
The property of the flow equation is explored in detail 
in Ref.\ \cite{FMS}, 
and a prescription for calculating the holographic Weyl anomaly 
is given for arbitrary dimensions. 
In this short note, we study possible ambiguities in the 
prescription that may arise from local terms in the flow equation, 
and point out that they actually do not yield 
physically meaningful effects to the Weyl anomaly. 
This result is not new and may have been noticed for experts. 
However, we believe that it is instructive to discuss it in some detail 
because this point can be clarified in a simple manner 
by the formulation given in Ref.\ \cite{FMS}. 
We discuss bulk gravity with scalar fields. 
The extension to other cases should be straightforward.

We start with the bulk action for a $(d+1)$-dimensional manifold 
$M_{d+1}$:
\ba
 S_{d+1}[\bG_{M\!N}(x,r),\phi^i(x,r)]
  \!\!&=&\!\!\int_{M_{d+1}}d^{d+1}\!X \sqrt{\bG}\left( 
  V(\phi)-\bR
  +{1\over 2}\,L_{ij}(\phi)\,\bG^{M\!N}\,\partial_M\phi^i\,
  \partial_N\phi^j 
  \right)\nn
 &&~-2\int_{\Sigma_d}d^dx\,\sqrt{G}\,K, 
 \label{bulk}
\ea
where $X^M=(x^\mu,r)$ with $\mu,\nu=1,2,\cdots,d$. 
The Euclidean time $r$ is regarded as the flow parameter of the RG, 
and we have introduced a UV cut-off $r_0$ such that $r_0\leq r<\infty$. 
The second term in eq.\ \eq{bulk} is the contribution from the boundary 
$\Sigma_d\equiv\partial M_{d+1}$ at $r=r_0$, 
which needs to be introduced in order for Dirichlet boundary 
conditions to be imposed consistently \cite{GH}. 
In what follows, we take the bulk metric $\bG_{M\!N}$ 
to be in the temporal gauge \cite{dVV}
\ba
 ds^2=\bG_{M\!N}\,dX^M dX^N
  =dr^2+G_{\mu\nu}(x,r)dx^{\mu}dx^{\nu}, 
\ea
for which the extrinsic curvature is given as 
$K=(1/2)\,G^{\mu\nu}\,\partial_r G_{\mu\nu}$.  
Let $\Gb_{\mu\nu}(x,r;G(x),r_0)$ and $\phib^i(x,r;\phi(x),r_0)$
be the classical trajectory of $G_{\mu\nu}(x,r)$ 
and $\phi^i(x,r)$ with the boundary condition 
\begin{equation}
 \Gb_{\mu\nu}(x,r\!=\!r_0)=G_{\mu\nu}(x),\qquad
 \phib^i(x,r\!=\!r_0)=\phi^i(x).
\end{equation}
We assume that the classical trajectory is uniquely 
determined by this initial value, 
demanding the regular behavior inside $M_{d+1}~(r\rightarrow+\infty)$.
The on-shell action $S[G(x),\phi(x)]$ is then defined 
as a functional of the boundary values 
and obtained by substituting the classical solution 
into the bulk action
\ba
 S[G_{\mu\nu}(x),\phi(x)] \equiv
 S_{d+1}\left[\,\Gb_{\mu\nu}(x,r;G(x),r_0),\,
  \phib^i(x,r;\phi(x),r_0)\,\right] .
\ea
Reflecting general covariance along the $r$-direction, 
the on-shell action does not depend on the coordinate value 
of the boundary, 
$r_0$, and obeys the {\em flow equation} \cite{dVV} 
\ba
 \Bigl\{ S,S\Bigr\}(x)=\sqrt{G(x)}\,\cL_d(x),
 \label{eq1}
\ea
where 
\ba
 \cL_d(x)\equiv V(\phi)-R+\frac{1}{2}\,L_{ij}(\phi)\,
  G^{\mu\nu}\,\partial_\mu\phi^i\,\partial_\nu\phi^j,
 \label{eq2}
\ea
and for arbitrary functionals $A$ and $B$, 
$\bigl\{A,B\bigr\}(x)$ is defined by
\begin{equation}
 \Bigl\{A,B\Bigr\}(x)\equiv{1\over\sqrt{G}}\left[
  \left( -{1\over d-1}\,G_{\mu\nu}G_{\rho\sigma}
    +G_{\mu\rho}G_{\nu\sigma}\right)
  {\delta A\over \delta G_{\mu\nu}}\,
   {\delta B\over \delta G_{\rho\sigma}}
   +{1\over 2}\,L^{ij}(\phi)\,
   {\delta A\over \delta\phi^i}\,{\delta B\over \delta\phi^j}
 \right].
 \label{eq3}
\end{equation}

To solve the flow equation \eq{eq1}--\eq{eq3}, 
we first decompose the on-shell action as 
\ba
 S[G(x),\phi(x)]= S_{\rm loc}[G(x),\phi(x)]+\Gamma[G(x),\phi(x)], 
\ea
where $S_{\rm loc}$ is the local part that can be written 
as an integral of a local function, 
and $\Gamma$ gives the generating functional of the boundary field
theory \cite{dVV}. 
Note that there is an ambiguity in this decomposition, 
reflecting that one can add any local terms to $\Gamma$.  
However, by introducing the following weight $w$ 
and by assigning the vanishing weight to $\Gamma$ \cite{FMS}, 
one can make this decomposition unique up to additions of 
local terms of weight 0: 
\begin{center}
\begin{tabular}{c|c}
       & weight $w$ \\ \hline
$G_{\mu\nu}(x), \,\phi^i(x), \,\Gamma[G,\phi]$ & $0$ \\ \hline
$\partial_{\mu}$ & $1$ \\ \hline
$R, \,R_{\mu\nu},\, \partial_\mu\phi^i\partial_\nu\phi^j,\,
\cdots$ & $2$ \\ \hline
$\delta \Gamma / \delta G_{\mu\nu}(x),\, 
\delta \Gamma / \delta \phi^i(x)$ & $d$ 
\end{tabular}
\end{center}
$\Sloc$ is then expanded with respect to this weight: 
\begin{equation}
 S_{\rm loc}[G(x),\phi(x)]=\int d^dx\sqrt{G(x)}\sum_{w=0,2,4,\cdots}
 \left[ {\cal L}_{\rm loc}(x)\right]_w.
\end{equation}
The first few terms can be parametrized as 
\begin{eqnarray}
[{\cal L}_{\rm loc}]_0=W(\phi),\qquad
 [{\cal L}_{\rm loc}]_2=-\Phi(\phi)\,R
  +{1\over 2}\,M_{ij}(\phi)\,G^{\mu\nu}\partial_{\mu}\phi^i\,
  \partial_{\nu}\phi^j.
\end{eqnarray}
One can show that $\bigl[\cLloc\bigr]_0,\cdots,\bigl[\cLloc\bigr]_{d-2}$ 
are determined by the flow equation at weight $w=0,\cdots,d-2$, 
and can be written in terms of $V(\phi)$ and $L_{ij}(\phi)$ 
\cite{dVV}\cite{FMS}. 
On the other hand, the flow equation at weight $w=d$ 
gives the following equation that would determine $\Gamma$:
\begin{equation}
 -2\,G_{\mu\nu}\,{\delta\Gamma\over \delta G_{\mu\nu}}
  +\beta^i\,{\delta\Gamma\over\delta\phi^i}
  =-\,{1\over [\gamma]_0}\,\Bigl[ \{ S_{\rm loc},S_{\rm loc}\}\Bigr]_d, 
\label{flow;eq}
\end{equation}
where
\begin{equation}
 [\gamma]_0={W(\phi)\over 2(d-1)},\qquad
  \beta^i={2(d-1)\over W(\phi)}\,L^{ij}(\phi)\,\partial_jW(\phi). 
\end{equation}
The right-hand side of eq.\ (\ref{flow;eq})  generally 
consists of the $d$-dimensional Weyl anomaly ${\cal W}_d$ 
of the boundary field theory and the contribution 
from the $\bigl[\cLloc\bigr]_d$ \cite{FMS}: 
\begin{equation}
 -\,{1\over [\gamma]_0}\,\Bigl[ \{ S_{\rm loc},S_{\rm loc}\}\Bigr]_d
  =-2\,\sqrt{G}\,{\cal W}_d
  -{2\over [\gamma]_0}\,\Bigl\{ S_{{\rm loc};\,-d},S_{{\rm loc};\,0} \Bigr\},
\end{equation}
where we have introduced
\begin{equation}
S_{{\rm loc};\,\,w-d}\equiv\int d^dx\,\sqrt{G(x)}\,[{\cal L}_{\rm loc}]_w. 
\end{equation}
The weight shifts by $-d$ after the integration 
because the weight of $d^d x$ is $-d$. 
Since the Weyl anomaly $\cW_d$ can be totally written 
in terms of $\bigl[\cLloc\bigr]_0,\cdots,\bigl[\cLloc\bigr]_{d-2}$, 
eq.\ \eq{flow;eq} shows that $\Gamma$ can only be determined 
up to contributions from $\bigl[\cLloc\bigr]_d$. 
However, by using the relations 
\begin{equation}
 {\delta S_{{\rm loc};\,-d}\over\delta G_{\mu\nu}}=
 {\sqrt{G}\over 2}\,W(\phi)\,G^{\mu\nu},\qquad
 {\delta S_{{\rm loc};\,-d}\over\delta \phi^i}=
 \sqrt{G}\,\,\partial_i W(\phi), 
\end{equation}
one finds that 
\ba
 - \,{1\over [\gamma]_0}\Bigl[ \{ S_{\rm loc},S_{\rm loc}\}\Bigr]_d
  =-2\,\sqrt{G}\,{\cal W}_d
  +2\,G_{\mu\nu}{\delta S_{{\rm loc};\,0}\over \delta G_{\mu\nu}}
  -\beta^i\,{\delta S_{{\rm loc};\,0}\over \delta\phi^i}, 
 \label{slocd}
\ea
and can rewrite the flow equation (\ref{flow;eq}) as 
\begin{equation}
-2\,G_{\mu\nu}{\delta\over\delta G_{\mu\nu}}(\Gamma+S_{{\rm loc};\,0})
+\beta^i{\delta\over\delta \phi^i}(\Gamma+S_{{\rm loc};\,0})
=-2\,\sqrt{G}\,{\cal W}_d. 
\end{equation}
Thus, we have seen that the contribution from the term 
$\bigl[\cLloc\bigr]_d$ can be absorbed into $\Gamma$ 
by redefining it as $\Gamma'=\Gamma+S_{{\rm loc};\,0}$.
Note that $\Gamma'$ still has vanishing weight.

Instead of redefining $\Gamma$, one can modify the Weyl anomaly 
without making any essential change. 
To show this, we first notice that the second term in eq.\ 
\eq{slocd} can be written as a total derivative: 
\begin{equation}
 2\,G_{\mu\nu}\,{\delta S_{{\rm loc};\,0}\over\delta G_{\mu\nu}}
  =-2\,\sqrt{G}\,\nabla_{\mu}{\cal J}^{\mu}_d 
 \label{totalder}
\end{equation}
with ${\cal J}_d^{\mu}$ some local current. 
In fact, for infinitesimal Weyl transformations 
($\sigma(x)\ll 1$: arbitrary function), 
we have  
\begin{equation}
 S_{{\rm loc};\,0}[e^{\sigma (x)}G(x),\phi(x)]
  -S_{{\rm loc};\,0}[G(x),\phi(x)]
  =\int d^dx\,\sigma (x)\,
  G_{\mu\nu}\,{\delta S_{{\rm loc};0}\over\delta G_{\mu\nu}}.
 \label{dilatation}
\end{equation}
One can easily understand that 
$S_{{\rm loc};\,0}[G(x),\phi(x)]$ is invariant 
under {\em constant} Weyl transformations 
($G_{\mu\nu}(x)\!\rightarrow\!e^\sigma G_{\mu\nu}(x)$,
$\phi^i(x)\!\rightarrow\!\phi^i(x)$ with  $\sigma$ constant), 
so that the left-hand side of eq.\ \eq{dilatation} 
can generally be written as
\ba
  \int d^d x \,\partial_\mu\sigma(x)\,\sqrt{G}\,\cJ_d^\mu
\ea
with some local function $\cJ_d^\mu$. 
By integrating this by parts and comparing the result with the 
right-hand side of eq.\ \eq{dilatation}, 
one obtains eq.\ \eq{totalder}. 
Thus we have shown that eq.\ \eq{flow;eq} can be rewritten 
into the following form: 
\begin{equation}
 -2\,G_{\mu\nu}\,{\delta\Gamma\over\delta G_{\mu\nu}}
  +\beta^i\,{\delta\Gamma\over\delta\phi^i}
  =-2\,\sqrt{G}\,({\cal W}_d+\nabla_{\mu}{\cal J}^{\mu}_d)
  -\beta^i\,{\delta S_{{\rm loc};\,0}\over \delta\phi^i}. 
\end{equation}
This implies that when we take $\Gamma$ as the generating functional, 
the Weyl anomaly differs from $\cW_d$ 
only by a total derivative at conformal fixed points ($\beta^i=0$).

In summary, in the formulation of the holographic RG 
based on the flow equation, 
the ambiguity in defining the generating functional 
$\Gamma$ totally corresponds (as expected) to 
the addition of a total derivative term 
to the Weyl anomaly when the beta functions vanish.

\noindent{\bf Acknowledgment}

The authors would like to thank S.\ Matsuura for a useful discussion. 
The work of M.F. is supported in part by a Grand-in-Aid
for Scientific Research from the Ministry of Education, Science,
Sports and Culture,
and the work of T.S.\ is supported in part
by JSPS Research Fellowships for Young Scientists.


\begin{thebibliography}{99}

\bibitem{MGKPW}
J.\ Maldacena,
``{\it The large $N$ limit of superconformal field theories and 
supergravity},''
Adv.\ Theor.\ Math.\ Phys.\ {\bf 2} (1998) 231,
hep-th/9711200;\\
S.\ S.\ Gubser, I.\ R.\ Klebanov and A.\ M.\ Polyakov,
``{\it Gauge Theory Correlators from Non-Critical String Theory},''
Phys.\ Lett.\ {\bf B428} (1998) 105,
hep-th/9802109;\\
E.\ Witten,
``{\it Anti De Sitter Space And Holography},''
Adv.\ Theor.\ Math.\ Phys.\ {\bf 2} (1998) 253,
hep-th/9802150.

\bibitem{review}
O.\ Aharony, S.\ S.\ Gubser, J.\ Maldacena, H.\ Ooguri and Y.\ Oz,
``{\it Large N Field Theories, String Theory and Gravity},''
hep-th/9905111, and references therein.

\bibitem{E.T.A}
E.\ T.\ Akhmedov,
``{\it A remark on the AdS/CFT correspondence and the renormalization
 group flow},''
Phys.Lett. {\bf B442} (1998) 152,
hep-th/9806217.

\bibitem{AG}
E.\ Alvarez and C.\ Gomez,
``{\it Geometric Holography, the Renormalization Group 
and the c-Theorem},''
Nucl.Phys. {\bf B541} (1999) 441,
hep-th/9807226.

\bibitem{HRG}
D.Z.\ Freedman, S.S.\ Gubser, K. Pilch and N.P.\ Warner, 
``{\it Renormalization Group Flows from Holography--Supersymmetry 
and a c-Theorem},''
hep-th/9904017.

\bibitem{GPPZ}
L.\ Girardello, M.\ Petrini, M.\ Porrati and A.\ Zaffaroni, 
``{\it Novel Local CFT and Exact Results on Perturbations of N=4 Super
	Yang Mills from AdS Dynamics},''
JHEP {\bf 12} (1998) 022,
hep-th/9810126.

\bibitem{GPPZ-2}
L.\ Girardello, M.\ Petrini, M.\ Porrati and A.\ Zaffaroni
``{\it The Supergravity Dual of N=1 Super Yang-Mills Theory},''
Nucl.Phys. {\bf B569} (2000) 451,
hep-th/9909047.

\bibitem{PS}
M.\ Porrati and A.\ Starinets,
{\it  ``RG Fixed Points in Supergravity Duals of 4-d Field Theory and
	Asymptotically AdS Spaces},''
Phys.Lett. {\bf B454} (1999) 77,
hep-th/9903085.

\bibitem{BPPZ}
V.\ Balasubramanian and P.\ Kraus,
``{\it Spacetime and the Holographic Renormalization Group},''
Phys.Rev.Lett. {\bf 83} (1999) 3605,
hep-th/9903190.

\bibitem{ST}
K.\ Skenderis and P.\ K.\ Townsend,
{\it  ``Gravitational Stability and Renormalization-Group Flow},''
Phys.Lett. {\bf B468} (1999) 46,
hep-th/9909070.

\bibitem{DFGK}
O.\ DeWolfe, D.\ Z.\ Freedman, S.\ S.\ Gubser and A.\ Karch,
``{\it Modeling the fifth dimension with scalars and gravity},''
hep-th/9909134.

\bibitem{dVV}
J.\ de Boer, E.\ Verlinde and H.\ Verlinde,
``{\it On the Holographic Renormalization Group},''
hep-th/9912012.

\bibitem{HS;weyl}
M.\ Henningson and K.\ Skenderis,
``{\it The Holographic Weyl anomaly},''
JHEP {\bf 07} (1998) 023,
hep-th/9806087.

\bibitem{FMS}
M.\ Fukuma, S.\ Matsuura and T.\ Sakai, 
``{\it A Note on the Weyl Anomaly in the Holographic 
Renormalization Group},''
hep-th/0007062.

\bibitem{GH}
G.\ W.\ Gibbons and S.\ W.\ Hawking,
``{\it Action Integrals and Partition Functions in Quantum Gravity},''
Phys. Rev. {\bf D15} (1977) 2752.


\end{thebibliography}
\end{document}